\documentclass[twocolumn]{aastex62}

\usepackage{verbatim,float,xcolor,graphicx}

\newcommand{\zwindow}{$0.3<z<0.4$}
\newcommand{\msun}{$M_{\odot}$}
\newcommand{\massrangelo}{$10^{9.0} \leq M/M_{\odot} \leq 10^{9.5}$}
\newcommand{\massrange}{$10^{9} \leq M/M_{\odot} \leq 10^{10}$}
\newcommand{\sigmacut}{$\Sigma_5<10^{0.3}$~Mpc$^{-2}$}
\newcommand{\donecut}{$D_1>10^{-0.7}$~Mpc}
\newcommand{\niso}{322} 
\newcommand{\sfr}{$M_{\odot}~$yr$^{-1}$}
\newcommand{\ssfr}{yr$^{-1}$}

\newcommand{\plotdir}{}

\shortauthors{Patel et al.}

\begin{document}

\title{Testing the Breathing Mode in Intermediate Mass Galaxies and its Predicted Star Formation Rate-Size Anti-Correlation\footnote{Based on observations made with the NASA/ESA Hubble Space Telescope, obtained at the Space Telescope Science Institute, which is operated by the Association of Universities for Research in Astronomy, Inc., under NASA contract NAS 5-26555. These observations are associated with programs 9822 and 10092.}}

\correspondingauthor{Shannon G. Patel}
\email{patel@carnegiescience.edu}

\author{Shannon G. Patel}
\affil{Carnegie Observatories, 813 Santa Barbara Street, Pasadena, CA 91101, USA}

\author{Daniel D. Kelson}
\affil{Carnegie Observatories, 813 Santa Barbara Street, Pasadena, CA 91101, USA}

\author{Nicholas Diao}
\affiliation{Carnegie Observatories, 813 Santa Barbara Street, Pasadena, CA 91101, USA}
\affiliation{Pomona College, Claremont, CA 91711, USA}

\author{Stephanie Tonnesen}
\affiliation{Center for Computational Astrophysics, Flatiron Institute, 162 5th Ave., New York, NY 10010, USA}

\author{Louis E. Abramson}
\affiliation{Department of Physics \& Astronomy, UCLA, 430 Portola Plaza, Los Angeles, CA 90095-1547, USA}

\begin{abstract}

Recent hydrodynamical simulations predict that stellar feedback in intermediate mass galaxies (IMGs) can drive strong fluctuations in structure (e.g., half-light radius, $R_e$).  This process operates on timescales of only a few hundred Myr and persists even at late cosmic times.  One prediction of this quasi-periodic, galactic-scale ``breathing'' is an {\em anti}-correlation between star formation rate (SFR) and half-light radius as central gas overdensities lead to starbursts whose feedback drags stars to larger radii while star formation dwindles.  We test this prediction with a sample of \niso\ {\em isolated} IMGs with stellar masses of \massrangelo\ at \zwindow\ in the HST $I_{814}$ COSMOS footprint.  We find that IMGs with higher specific SFRs (SSFR~$>10^{-10}$~\ssfr) are the most extended with median sizes of $R_e \sim 3-3.4$~kpc and are mostly disk-dominated systems.  In contrast, IMGs with lower SSFRs ($<10^{-10}$~\ssfr) are a factor of $\sim 2-3$ more compact with median sizes of $R_e \sim 0.9-1.6$~kpc and have more significant bulge contributions to their light.  These observed trends are opposite the predictions for stellar feedback that operate via the ``breathing''  process described above.  We discuss various paths to reconcile the observations and simulations, all of which likely require a different implementation of stellar feedback in IMGs that drastically changes their predicted formation history.

\end{abstract}

\keywords{galaxies: evolution -- galaxies: formation -- galaxies: structure -- galaxies: star formation}

% ############################################################
\section{Introduction} \label{sec_intro}

Feedback from stars is predicted to play a critical role in galaxy formation, from regulating the conversion of gas into stars and shaping star formation histories \citep[SFHs,][]{gerola1980,stinson2007,hopkins2014}, to depositing metals into the CGM and IGM via galactic winds \citep{larson1974c,dekel1986,oppenheimer2006}, and even sculpting the morphologies and density profiles of galaxies \citep[][]{governato2010,chan2015,elbadry2016}.  These predictions are motivated and constrained by various observations, such as the slope of the faint end of the galaxy stellar mass function \citep{cole2001,oppenheimer2010}, the detection of galactic-scale outflows among star forming galaxies \citep[SFGs,][]{rubin2014}, and the presence of cored central density profiles \citep{marchesini2002,simon2005,deblok2008}.

Due to limited resolution and the challenges of modeling baryonic physics, most hydrodynamical simulations implement stellar feedback using ``sub-grid'' prescriptions tuned or calibrated to match various observations \citep[e.g.,][]{vogelsberger2014b,schaye2015}.  However, the FIRE simulation has taken an important step by implementing explicit stellar feedback models \citep{hopkins2014}.  As a result, their simulations may be more predictive of the evolution of galaxies across a range of halo masses.  Indeed, one of the successes of their simulation is the ability to approximate various integrated observations using recipes based on small scales \citep{hopkins2011c,hopkins2012c}, such as the stellar mass-halo mass relation, the star forming sequence (SFS), and late-time ($z \lesssim 1$) stellar mass growth for subsets of the galaxy population.

FIRE also makes predictions for the evolution in galaxy structure, which for galaxies with stellar masses of $10^{7}\lesssim M/M_{\odot} \lesssim 10^{9.6}$, appears to be strongly coupled to star formation activity and therefore stellar feedback \citep{chan2015,elbadry2016}.  Specifically, in FIRE, gas cools and fuels a central starburst, which is subsequently curtailed by strong stellar feedback.  As star formation subsides and the gas is expelled to large radii, it drives fluctuations in the gravitational potential causing the stars to migrate outward as well.  Eventually the gas cools and falls back to the center, reversing the process and fueling another episode of concentrated star formation.  This rapid, bursty cycle in FIRE \citep[][]{sparre2017,anglesalcazar2017b} continues even at late cosmic times.  One key observational prediction of this implementation of stellar feedback is that intermediate mass galaxies (IMGs) undergo quasi-periodic fluctuations in SSFR, which is {\em anti}-correlated with half-light radius.

In this work, we test that prediction against an observed sample of IMGs.  Assembling such a sample is not trivial.  There are only four galaxies near the Local Group spanning stellar masses \massrange: the LMC, M33, NGC 55, and NGC 300 \citep{mcconnachie2012}.  In addition, their formation histories and properties may not be cosmologically representative due to their proximity to more massive galaxies (i.e., the Milky Way and Andromeda).  For this study, we therefore select a large sample of {\em isolated} IMGs from the HST $I_{814}$ COSMOS footprint \citep{scoville2007b} at \zwindow.  Despite the higher redshift range, the HST imaging affords a factor of two boost in spatial resolution (in kpc) compared to ground based imaging at lower redshifts (e.g., SDSS).

We assume a cosmology with $H_0=70$~km~s$^{-1}$~Mpc$^{-1}$, $\Omega_M=0.3$ and $\Omega_{\Lambda}=0.7$.  Stellar masses and SFRs are based on a \citet{chabrier2003} IMF.  All magnitudes are AB.

\section{Data and Analysis}

\begin{figure*}
\epsscale{1.1}
\plotone{\plotdir 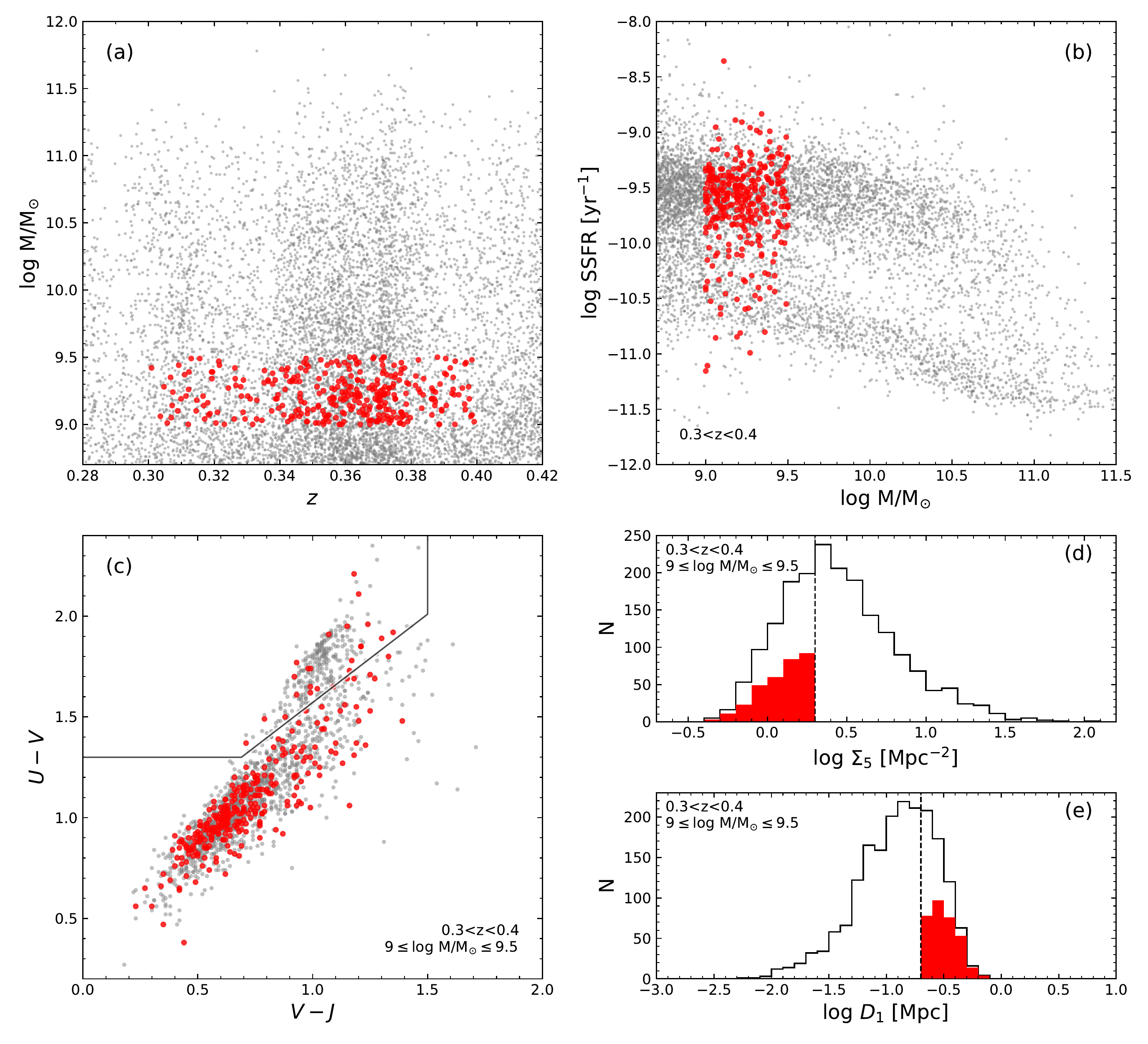}
\caption{Selection of \niso\ isolated IMGs at \zwindow\ with stellar mass \massrangelo\ in COSMOS.  (a) Stellar mass vs. redshift. (b) SSFR vs. stellar mass for galaxies at \zwindow. (c) Rest-frame $U-V$ vs. $V-J$ colors.  Environmental distributions of (d) projected local galaxy density ($\Sigma_5$) and (e) distance to nearest neighbor ($D_1$).  In all panels, red denotes isolated galaxies with ACS $I_{814}$ imaging, where isolated is defined as $\Sigma_5<10^{0.3}$~Mpc$^{-2}$ and $D_1>10^{-0.7}$~Mpc (vertical dashed lines).  The union of these criteria is represented by the red histograms.} \label{fig_selection}
\end{figure*}

\begin{figure*}
\epsscale{1.2}
\plotone{\plotdir 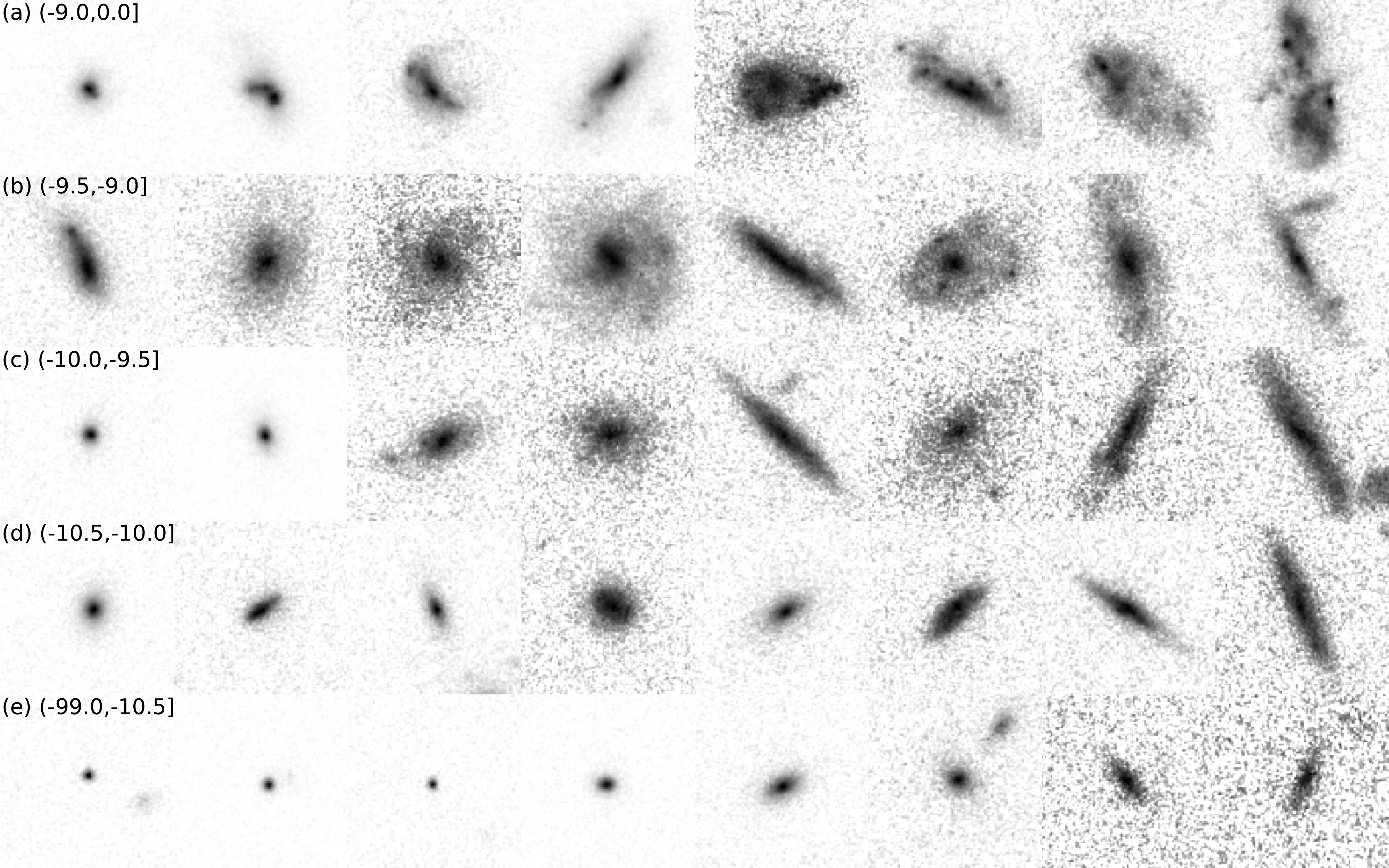}
\caption{Example HST ACS $I_{814}$ postage stamps of randomly selected, isolated IMGs with stellar mass \massrangelo\  at \zwindow\ in COSMOS.  Each row is a bin in SSFR and is further sorted by half-light radius.  An inverse sinh stretch is employed to reveal lower surface brightness features along with bright cores.  Each stamp is $\sim 3\farcs4$ on a side ($\sim 15-18$~kpc).  IMGs with higher SSFRs are typically larger in size compared to those with lower SSFRs. } \label{fig_ps}
\end{figure*}

\begin{figure*}
\epsscale{1.1}
\plottwo{\plotdir 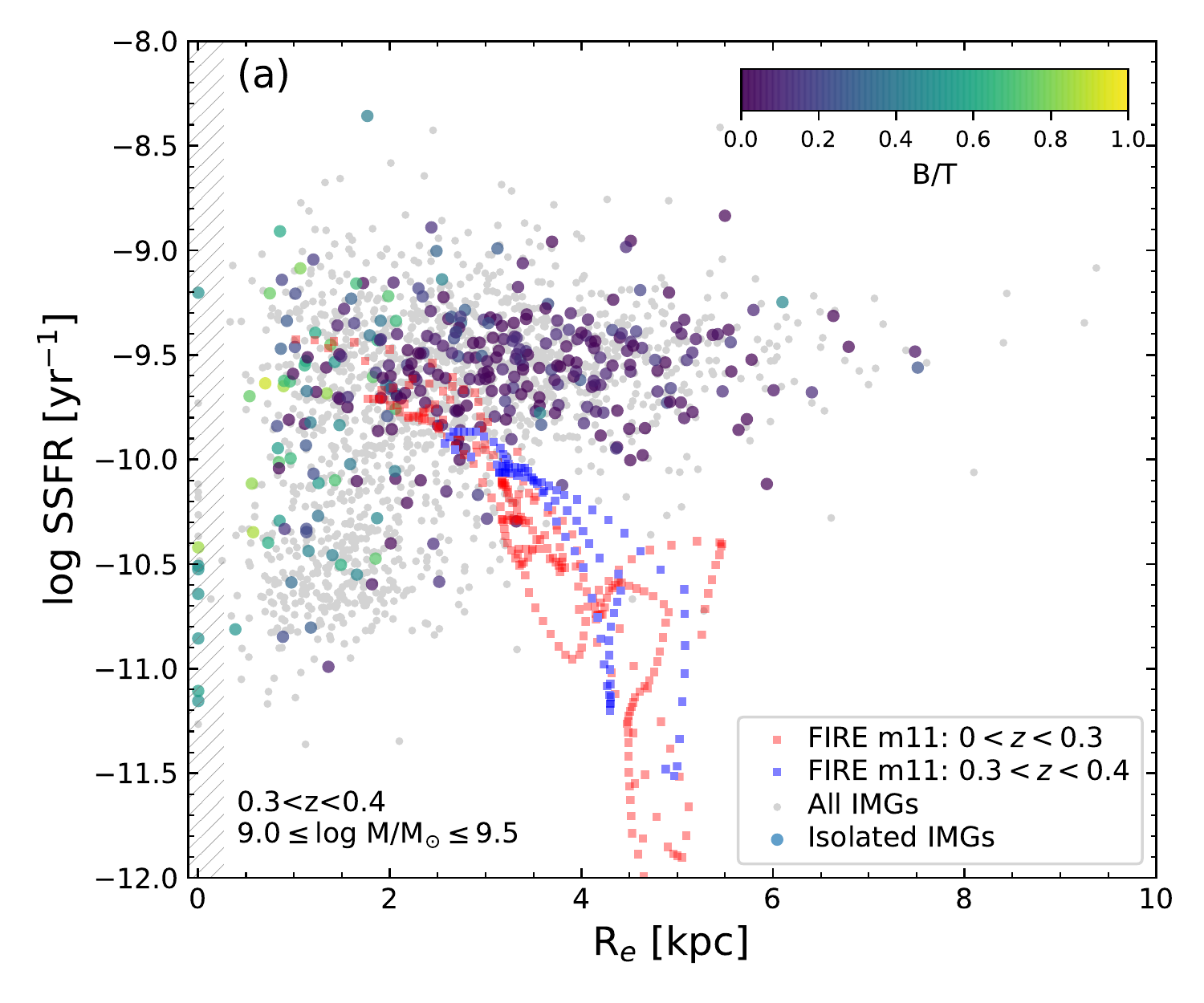}{\plotdir 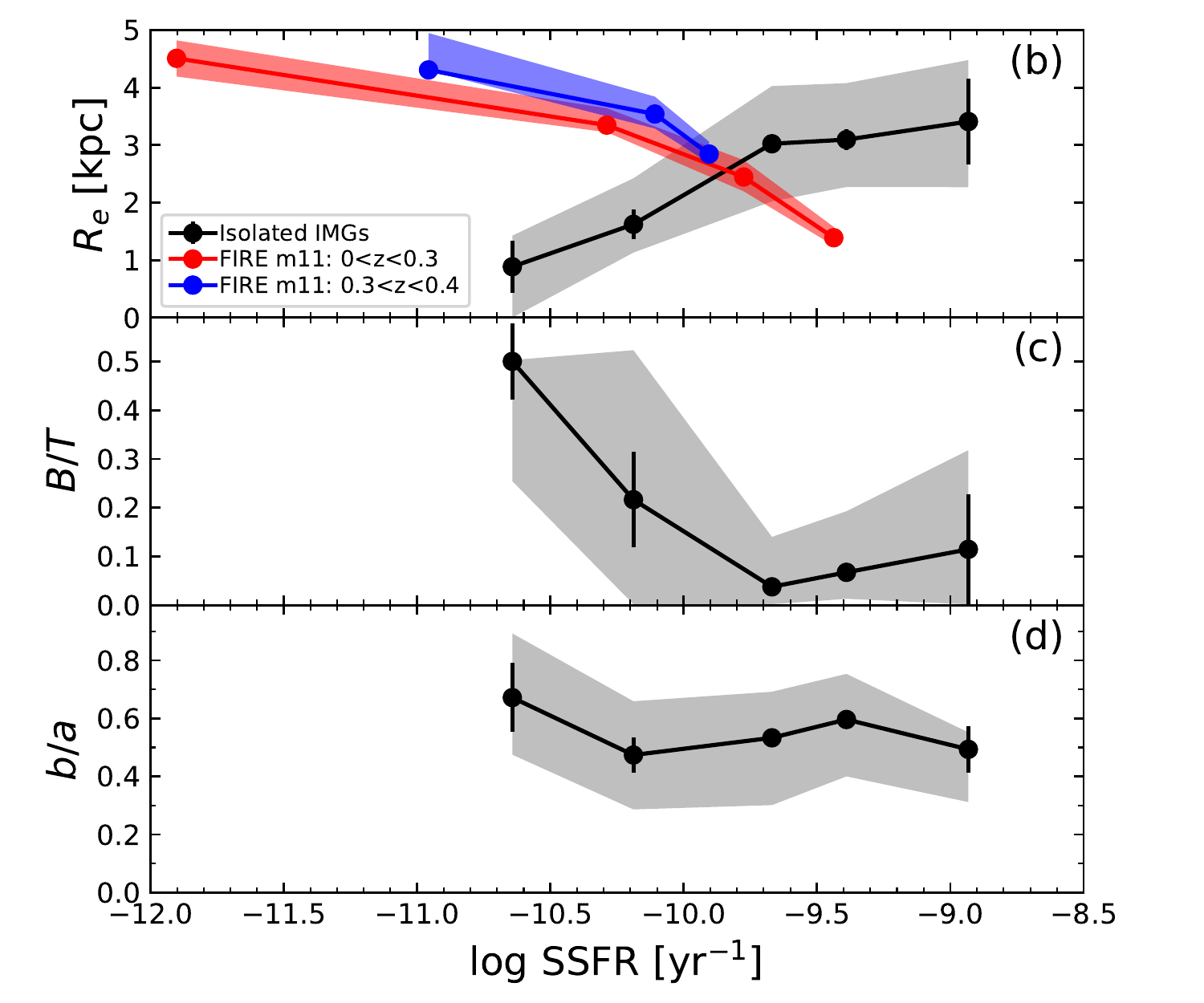}
\caption{(a) SSFR vs. half-light radius for IMGs with stellar mass \massrangelo\ at \zwindow\ (gray circles).  The sub-sample of isolated IMGs are color-coded according to their $B/T$.  The hatched region represents the $I_{814}$ PSF HWHM.  The squares shows the evolutionary track for the simulated galaxy m11 from \citet{elbadry2016} over $0<z<0.4$, sampled at timesteps of $10$~Myr.  The blue squares are the segment between \zwindow.  Median (b) $R_e$, (c) $B/T$, and (d) $b/a$ with $1\sigma$ bootstrapped errors in bins of SSFR for the isolated IMG sample (black) as well as m11 (blue and red; $R_e$ only).  The shaded regions indicate the interquartile range.  IMGs with SSFRs above $>10^{-10}$~\ssfr\ are generally $\sim 2-3$ times as extended as IMGs with lower SSFRs and also have lower $B/T$ values.  There is a significant offset between m11 and the observations, and more importantly, the anti-correlation between SSFR and $R_e$ in the models conflicts with the data. } \label{fig_ssfr_size}
\end{figure*}

We select IMGs from the UltraVISTA catalog of \citet{muzzin2013b}, which covers the HST ACS $I_{814}$ footprint in COSMOS.  In this work, we study IMGs at \zwindow\ with stellar masses of \massrangelo\ since these are the highest mass galaxies for which \citet{elbadry2016} predict stellar feedback to play a prominent role in shaping structure.  A small percentage of galaxies in the catalog have spectroscopic redshifts, which are used in place of the photometric redshift when available.  The photometric redshift uncertainty for our sample is $\sigma_z/(1+z) \approx 0.009$.  This is adequate for studying galaxies in different environments \citep[e.g.,][]{patel2010}.  The stellar mass completeness limit at $z=0.4$ is $M \sim 10^{8.7}$~\msun\ \citep{muzzin2013c}, which is a factor of two below the lowest masses in our sample.

We use the SFRs measured by \citet{muzzin2013b}, which are derived from a sum of the UV and IR flux \citep[e.g.,][]{bell2005}.  The MIPS 24~\micron\ $3\sigma$ detection limit at $z=0.4$ corresponds to a SFR of $\sim 0.45$~\sfr.  Most galaxies in the mass range studied here lack an IR detection.  However, stacking the MIPS imaging at the location of spatially isolated non-detections indicates that the median IR contribution at these low masses is small ($\sim 20\%$).  An analysis of lower redshift IR-detected IMGs confirms that the typical IR contribution to the total SFR is on this level.  We take the total SFR to be the sum of the UV and IR when the latter has a $>3\sigma$ detection, otherwise the SFR is based solely on the UV.  

The sample of IMGs selected on the criteria above, as well as on their environmental properties (see below), are shown as red circles in Figure~\ref{fig_selection} in various projections.  Panel ($a$) shows that the redshifts are precise enough to resolve the presence of large scale structure around $z \sim 0.37$.  Panel ($b$) shows that while most of the IMGs lie on the SFS, their SSFRs span a wide range, from starburst to quiescent.  The rest-frame $U-V$ vs. $V-J$ colors in panel ($c$) indicate that $\sim 84\%$ of the parent IMG sample lies in the star forming region of the diagram \citep[e.g.,][]{williams2009,patel2012}.

Here, we study IMGs in isolated environments so that we can compare to simulated galaxies in similar environments.  We compute two different environmental measures to select galaxies, (1) in under-dense regions and, (2) isolated from any single neighbor.  For (1), we compute the local galaxy density using the projected distance to the 5th nearest neighbor ($\Sigma_5$) with stellar mass $M>10^{10.3}$~\msun\ and within a redshift window of $|\Delta z|/(1+z)<0.02$.  For (2), we compute the projected distance to the nearest galaxy ($D_1$) with stellar mass $M>10^{9}$~\msun\ within the same redshift window as (1).  Galaxies near the survey edge with biased environmental measurements were excluded from both the parent sample and isolated sub-sample.  Figure~\ref{fig_selection} shows the distribution of $\Sigma_5$ and $D_1$ for IMGs with stellar mass \massrangelo.  We retain galaxies with \sigmacut\ and \donecut, resulting in $17\%$ of the parent sample designated as isolated (red circles in Figure~\ref{fig_selection}).  A higher proportion are SFGs ($\sim 93\%$) compared to the parent sample, given that the quenched satellite contribution is minimized by our selection.

We use the HST ACS $I_{814}$ imaging in COSMOS \citep[rest-frame $r$-band,][]{scoville2007b,koekemoer2007} in order to measure structural properties with one and two-component GALFIT Sersic profile fits \citep{peng2002}, incorporating similar techniques as discussed in \citet{patel2017}.  The one-component fits are used to determine the half-light radii, which for our sample agree well with the literature \citep[e.g.,][]{vanderwel2014}.  The two-component fits employ $n=1$ and $n=4$ Sersic profiles to represent disk and bulge components, respectively, allowing us to compute bulge-to-total ratios ($B/T$).  This quantity is an indicator of concentrated light (e.g., from a central starburst or an old spheroid).  We have carried out an exhaustive Monte Carlo analysis to verify the validity of the bulge-disk decompositions (Patel~et~al., in preparation).

\section{Results}

In Figure~\ref{fig_ps} we show example HST $I_{814}$ postage stamps of \massrangelo\ isolated IMGs at \zwindow.  Given the redshift range, the $\sim 3\farcs4$ postage stamps are $\sim 15-18$~kpc on a side.  Each row represents a bin in SSFR, with the highest star forming objects at the top, and the lowest on the bottom.  Galaxies are further sorted by their half-light radii within each SSFR bin.  The difference in structure between different SSFR bins is discernible.  Starbursting IMGs (SSFR~$>10^{-9}$~yr$^{-1}$), with stellar mass doubling times of $\lesssim 1$~Gyr (assuming sustained, constant star formation), appear somewhat diverse in their structure.  Most appear to harbor disks, some are compact, and a select few look as if they are part of a merger or contain several sites of intense star formation.  Meanwhile, IMGs that lie on the SFS ($10^{-10}<$~SSFR/yr$^{-1}$~$<10^{-9}$) mostly resemble extended disk-like galaxies, though a small few appear to be somewhat more compact and spheroidal.  Just below the bulk of the SFS ($10^{-10.5}<$~SSFR/yr$^{-1}$~$<10^{-10}$), where stellar mass doubling times exceed a Hubble time, the population is comprised predominantly of spheroids and the occasional disk.  Even further below this region, (SSFR~$<10^{-10.5}$~yr$^{-1}$), where galaxies are deemed quiescent by some metrics (e.g., SSFR~$<0.3/t_H$), almost all IMGs are compact spheroids.

We quantify these findings in Figure~\ref{fig_ssfr_size} and examine the connection between star formation and structural properties.  The SSFRs for the full IMG sample are plotted against their half-light radii in gray in panel ($a$).  The sub-sample of isolated IMGs are color-coded according to their $B/T$ ratios.  The median half-light radii for the different SSFR bins designated in Figure~\ref{fig_ps} are computed for isolated IMGs in panel ($b$).  As alluded to in Figure~\ref{fig_ps}, the most active star formers, have the largest median size of $R_e \sim 3.4$~kpc.  IMGs on the SFS have a range of sizes with a median of $3.0-3.1$~kpc.  Just below the SFS, the median size shrinks to $\sim 1.6$~kpc, and even further below (SSFR~$<10^{-10.5}$~yr$^{-1}$), the median size reaches a minimum of $\sim 0.9$~kpc.  This trend is in agreement with the findings of \citet{morishita2017}.  We note that applying a more stringent star-galaxy separation than delineated in \citet{muzzin2013b} would remove some of the unresolved objects within the PSF HWHM (hatched region in panel ($a$)), however, our conclusions would remain unchanged, even if all unresolved objects were removed.

Also shown in Figure~\ref{fig_ssfr_size} is the evolutionary track of the galaxy m11 (red and blue squares) from the FIRE simulations \citep{elbadry2016}, sampled at timesteps of 10~Myr.  The $z \sim 0$ morphology of the galaxy is described by \citet{hopkins2014} as a ``fluffy dwarf spheroidal''.  The displayed SFRs for this galaxy have been averaged over timescales of $\sim 100$~Myr, which is similar to the timescales probed by the UV+IR SFRs measured for our IMG sample.  The half-light radius for m11 is measured in the $r$-band, also the same as employed for our sample.  The simulated galaxy reaches a final mass at $z \sim 0$ of $M=10^{9.32}$~\msun.  Tracing its SFH back to $z \sim 0.4$, this galaxy remains within the stellar mass range studied in this work.  The rapid fluctuation in SFR causes the galaxy to oscillate between star forming and quiescent states.

The two-component Sersic fits reveal that the bulge component is generally more dominant among IMGs with low SSFRs (Figure~\ref{fig_ssfr_size}($c$)).  In that regime, the bulges are likely composed of older stellar populations as opposed to a concentrated starburst.  Figure~\ref{fig_ssfr_size}($d$) shows the median axis ratio, as well as the interquartile range.  About $\sim 43\%$ of isolated IMGs have $b/a<0.5$.  In contrast, \citet{elbadry2016} find that none of the FIRE galaxies have axis ratios below this threshold.  We note that restricting our sample to $b/a>0.5$ does not affect our conclusions.

We identify two glaring differences in comparing the model predictions to the observations: (1) there is a substantial offset in the SSFR-$R_e$ plane between m11 and the locus of observed isolated IMGs. (2) More importantly, the anti-correlation in the simulations between SSFR and $R_e$ is not seen in the observations.  We discuss these and other findings in the next section.

\section{Discussion and Conclusions}

\subsection{Comparing the simulation to the observations}

In the FIRE simulations, quasi-periodic stellar feedback is strongly coupled to fluctuations in the contemporaneous structure of present day galaxies with stellar masses $10^{7}\lesssim M/M_{\odot} \lesssim 10^{9.6}$, even at late times \citep{hopkins2014,elbadry2016}.  \citet{elbadry2016} find that order of magnitude changes in the SFR can lead to factor of two variations in the half-light radius for the simulated galaxy, m11, over only a few hundred Myr.  In fact, they argue these variations for single galaxies reproduce the bulk of the observed scatter in the size-mass relation at fixed stellar mass.  At higher masses ($M \gtrsim 10^{9.6}$~\msun), their simulations predict that deeper gravitational potentials are largely impervious to stellar feedback and therefore mitigate their effects, resulting in smoother star formation histories.  Meanwhile, at lower masses in the simulations ($M \lesssim 10^{7}$~\msun), star formation is less efficient and the resulting stellar feedback too weak to strongly impact the gravitational potential. 

Our findings in Figure~\ref{fig_ssfr_size} strongly challenge the FIRE predictions for the structural evolution of galaxies in the mass range studied here, \massrangelo.  As noted in the previous section, there is a substantial offset in the SSFR-$R_e$ parameter space spanned by the simulated galaxy, m11, and the observations.  The simulated galaxy is generally too large in size given its low SFR: while m11 spends $\sim 84\%$ of its time at \zwindow\ in the region with SSFR~$< 10^{-10}$~\ssfr\ and $R_e>3$~kpc, only $\sim 1-2\%$ of observed IMGs are found there.  In addition, most observed IMGs are SFGs, but m11 spends the vasty majority of its time below the bulk of the SFS. 

Even more consequential is the anti-correlation between SSFR and $R_e$ seen in the FIRE simulations \citep{elbadry2016}.  The Pearson correlation coefficient is $\sim -0.85$ for m11 over \zwindow\ and $\sim -0.63$ at $z<0.4$.  This anti-correlation forms the basis for one of FIRE's signature predictions, as described in Section~\ref{sec_intro}, of galaxies ``breathing''.  We do not see evidence for such a feedback loop in the observations if its presence indeed manifests in the form of an anti-correlation between SSFR and $R_e$.  Instead, the predicted and observed SSFR-$R_e$ distributions appear rather orthogonal.  One can also compare the example HST images presented here to those in Figure~2 of \citet{elbadry2016}.  Although the example in their figure is slightly lower mass ($M \sim 10^{8.5}$~\msun) than the IMG sample in this work, the relative comparison is the same: at their lowest SFRs, galaxies in the simulation are extended while the data show them to be compact.  Lastly, our data would also rule out a scenario where an IMG is terminally quenched by a single ``breath'' if the end result is an extended (e.g., $R_e \gtrsim 4$~kpc) QG.

\subsection{Caveats}

The primary caveat is that the comparison here is based on a small handful of simulated galaxies (or even a single galaxy in the case of m11).  However, given that all of the FIRE galaxies with $10^{7}<M/M_{\odot} \lesssim 10^{9.6}$ in \citet{elbadry2016} convey a similar narrative suggests that issues arising with their evolution appear to be a fated outcome of the simulation.  

It's possible that our environmental selection has not resulted in galaxies that are sufficiently isolated, but further constraining the environmental selection would shrink the sample far below the current $17\%$ of the parent IMG sample, at which point one must ask how representative and applicable any results are for understanding general IMG formation.  Moreover, it is highly unlikely that an SSFR-$R_e$ distribution similar to that of \citet{elbadry2016} could be extracted from the observations given the distribution of the parent population (gray circles in Figure~\ref{fig_ssfr_size}).  For example, employing a hypothetical environmental selection that returns $1\%$ of the parent population ($19$~galaxies), and selecting random sub-samples of 19 galaxies from the parent population, results in a negative Pearson correlation coefficient between SSFR and $R_e$ in $\sim 5\%$ of 10,000 trials and a value below $<-0.50$ only $0.03\%$ of the time.

While most of our IMGs at low SSFRs ($<10^{-10}$~\ssfr) lack an IR SFR component due to the MIPS detection limit, a stacking analysis at the location of the non-detections suggests low median obscured SFRs of $<0.09$~\sfr.  This is well below the $\sim 1.6$~\sfr\ required to shift the compact galaxies at SSFR~$<10^{-10}$~\ssfr\ to above $>10^{-9}$~\ssfr\ and invert the $R_e$-SSFR trend.

\subsection{Future Directions}

Our findings have several implications for numerical models of star formation and feedback.  While providing a detailed account is beyond the scope of this Letter, we briefly point to several avenues for potential reconciliation with observations.  (1) \citet{hopkins2014} point out that while varying numerical choices (e.g., SNe momentum coupling) have minimal impact on key integrated quantities such as stellar masses and SFHs, they can produce significant changes to galaxy structure.  (2) They also show that the non-linear combination of multiple feedback mechanisms (e.g., HII photoionization and radiation pressure, SNe, etc.) act together to drive the strong galactic winds that are responsible for the structural variations \citep[see also,][]{hopkins2012c,hopkins2012d}.  The degree to which these various mechanisms act in concert should therefore be explored further.  (3) In general, we find that m11 exhibits low SFRs compared to the data.  While this can be attributed to the galaxy lying somewhat below the KS-law \citep[Figure~8 in][]{hopkins2014}, addressing this issue would seemingly exacerbate the problem if stronger star formation in their simulation more strongly impacts structure.  (4) \citet{elbadry2016} find that more massive galaxies (e.g., m12i) do not exhibit a feedback loop between SSFR and $R_e$.  This is due to their deeper gravitational potential and because they have many more star forming regions that collectively overshadow the feedback from any single region.  This more gradual mode of star formation with less consequential stellar feedback may need to be ported to the lower masses studied here.  This is supported by observations that show bursty star formation to be more pronounced in galaxies mostly {\em below} $M<10^{9}$~\msun\ \citep{weisz2012}.  (5) The observed tail of low axis ratios suggests disks are more common at these masses than the simulations find.  Maintaining rotational support is therefore critical with any feedback implementation.  Collimated outflows could fulfill this requirement and are commonly found among SFGs near the IMG mass range \citep[e.g.,][]{chen2010,rubin2014}.

In upcoming work we will explore the properties of IMGs over a wider stellar mass range and in diverse environments.  The analysis will include recently obtained high S/N spectroscopic observations with IMACS on Magellan.  Our aim is to understand the diversity of formation histories of IMGs beyond the four in the vicinity of the Local Group.

\acknowledgments

We thank Andrew Benson for helpful discussions.


\begin{thebibliography}{38}
\expandafter\ifx\csname natexlab\endcsname\relax\def\natexlab#1{#1}\fi
\bibitem[{{Angl{\'e}s-Alc{\'a}zar} {et~al.}(2017)}]{anglesalcazar2017b}{Angl{\'e}s-Alc{\'a}zar}, D.,  {Faucher-Gigu{\`e}re}, C.-A.,  {Kere{\v s}}, D., {et~al.} 2017, \mnras, 470, 4698

\bibitem[{{Bell} {et~al.}(2005)}]{bell2005}{Bell}, E.~F.,  {Papovich}, C.,  {Wolf}, C., {et~al.} 2005, \apj, 625, 23

\bibitem[{{Chabrier}(2003)}]{chabrier2003}{Chabrier}, G.  2003, \pasp, 115, 763

\bibitem[{{Chan} {et~al.}(2015)}]{chan2015}{Chan}, T.~K.,  {Kere{\v s}}, D.,  {O{\~n}orbe}, J., {et~al.} 2015, \mnras, 454,2981

\bibitem[{{Chen} {et~al.}(2010)}]{chen2010}{Chen}, Y.-M.,  {Tremonti}, C.~A.,  {Heckman}, T.~M., {et~al.} 2010, \aj, 140, 445

\bibitem[{{Cole} {et~al.}(2001)}]{cole2001}{Cole}, S.,  {Norberg}, P.,  {Baugh}, C.~M., {et~al.} 2001, \mnras, 326, 255

\bibitem[{{de Blok} {et~al.}(2008)}]{deblok2008}{de Blok}, W.~J.~G.,  {Walter}, F.,  {Brinks}, E., {et~al.} 2008, \aj, 136, 2648

\bibitem[{{Dekel} \& {Silk}(1986)}]{dekel1986}{Dekel}, A. \, \&  {Silk}, J.  1986, \apj, 303, 39

\bibitem[{{El-Badry} {et~al.}(2016)}]{elbadry2016}{El-Badry}, K.,  {Wetzel}, A.,  {Geha}, M., {et~al.} 2016, \apj, 820, 131

\bibitem[{{Gerola} {et~al.}(1980)}]{gerola1980}{Gerola}, H., {Seiden}, P.~E., \&  {Schulman}, L.~S.  1980, \apj, 242, 517

\bibitem[{{Governato} {et~al.}(2010)}]{governato2010}{Governato}, F.,  {Brook}, C.,  {Mayer}, L., {et~al.} 2010, \nat, 463, 203

\bibitem[{{Hopkins} {et~al.}(2014)}]{hopkins2014}{Hopkins}, P.~F.,  {Kere{\v s}}, D.,  {O{\~n}orbe}, J., {et~al.} 2014, \mnras, 445,581

\bibitem[{{Hopkins} {et~al.}(2011)}]{hopkins2011c}{Hopkins}, P.~F., {Quataert}, E., \&  {Murray}, N.  2011, \mnras, 417, 950

\bibitem[{{Hopkins} {et~al.}(2012{\natexlab{a}})}]{hopkins2012d} {Hopkins}, P.~F., {Quataert}, E., \& {Murray}, N. 2012{\natexlab{a}}, \mnras, 421, 3522

\bibitem[{{Hopkins} {et~al.}(2012{\natexlab{b}})}]{hopkins2012c} {Hopkins}, P.~F., {Quataert}, E., \& {Murray}, N. 2012{\natexlab{b}}, \mnras, 421, 3488

\bibitem[{{Koekemoer} {et~al.}(2007)}]{koekemoer2007}{Koekemoer}, A.~M.,  {Aussel}, H.,  {Calzetti}, D., {et~al.} 2007, \apjs, 172, 196

\bibitem[{{Larson}(1974)}]{larson1974c}{Larson}, R.~B.  1974, \mnras, 169, 229

\bibitem[{{Marchesini} {et~al.}(2002)}]{marchesini2002}{Marchesini}, D.,  {D'Onghia}, E.,  {Chincarini}, G., {et~al.} 2002, \apj, 575, 801

\bibitem[{{McConnachie}(2012)}]{mcconnachie2012}{McConnachie}, A.~W.  2012, \aj, 144, 4

\bibitem[{{Morishita} {et~al.}(2017)}]{morishita2017}{Morishita}, T.,  {Abramson}, L.~E.,  {Treu}, T., {et~al.} 2017, \apj,835, 254

\bibitem[{{Muzzin} {et~al.}(2013{\natexlab{a}})}]{muzzin2013c}{Muzzin}, A.,  {Marchesini}, D.,  {Stefanon}, M., {et~al.} 2013{\natexlab{a}}, \apj, 777, 18

\bibitem[{{Muzzin} {et~al.}(2013{\natexlab{b}})}]{muzzin2013b}{Muzzin}, A.,  {Marchesini}, D.,  {Stefanon}, M., {et~al.} 2013{\natexlab{b}}, \apjs, 206, 8

\bibitem[{{Oppenheimer} \& {Dav{\'e}}(2006)}]{oppenheimer2006}{Oppenheimer}, B.~D. \, \&  {Dav{\'e}}, R.  2006, \mnras, 373, 1265

\bibitem[{{Oppenheimer} {et~al.}(2010)}]{oppenheimer2010}{Oppenheimer}, B.~D.,  {Dav{\'e}}, R.,  {Kere{\v s}}, D., {et~al.} 2010, \mnras, 406, 2325

\bibitem[{{Patel}(2010)}]{patel2010}{Patel}, S.  2010, PhD thesis, University of California, Santa Cruz

\bibitem[{{Patel} {et~al.}(2012)}]{patel2012}{Patel}, S.~G.,  {Holden}, B.~P.,  {Kelson}, D.~D., {et~al.} 2012, \apjl, 748, L27

\bibitem[{{Patel} {et~al.}(2017)}]{patel2017}{Patel}, S.~G., {Hong}, Y.~X., {Quadri}, R.~F., {Holden}, B.~P., \&  {Williams},R.~J.  2017, \apj, 839, 127

\bibitem[{{Peng} {et~al.}(2002)}]{peng2002}{Peng}, C.~Y., {Ho}, L.~C., {Impey}, C.~D., \&  {Rix}, H.-W.  2002, \aj, 124, 266

\bibitem[{{Rubin} {et~al.}(2014)}]{rubin2014}{Rubin}, K.~H.~R.,  {Prochaska}, J.~X.,  {Koo}, D.~C., {et~al.} 2014, \apj, 794, 156

\bibitem[{{Schaye} {et~al.}(2015)}]{schaye2015}{Schaye}, J.,  {Crain}, R.~A.,  {Bower}, R.~G., {et~al.} 2015, \mnras,446, 521

\bibitem[{{Scoville} {et~al.}(2007)}]{scoville2007b}{Scoville}, N.,  {Abraham}, R.~G.,  {Aussel}, H., {et~al.} 2007, \apjs, 172, 38

\bibitem[{{Simon} {et~al.}(2005)}]{simon2005}{Simon}, J.~D., {Bolatto}, A.~D., {Leroy}, A., {Blitz}, L., \&  {Gates}, E.~L. 2005, \apj, 621, 757

\bibitem[{{Sparre} {et~al.}(2017)}]{sparre2017}{Sparre}, M.,  {Hayward}, C.~C.,  {Feldmann}, R., {et~al.} 2017, \mnras, 466, 88

\bibitem[{{Stinson} {et~al.}(2007)}]{stinson2007}{Stinson}, G.~S., {Dalcanton}, J.~J., {Quinn}, T., {Kaufmann}, T., \& {Wadsley}, J.  2007, \apj, 667, 170

\bibitem[{{van der Wel} {et~al.}(2014)}]{vanderwel2014}{van der Wel}, A.,  {Franx}, M.,  {van Dokkum}, P.~G., {et~al.} 2014, \apj, 788, 28

\bibitem[{{Vogelsberger} {et~al.}(2014)}]{vogelsberger2014b}{Vogelsberger}, M.,  {Genel}, S.,  {Springel}, V., {et~al.} 2014, \mnras, 444,1518

\bibitem[{{Weisz} {et~al.}(2012)}]{weisz2012}{Weisz}, D.~R.,  {Johnson}, B.~D.,  {Johnson}, L.~C., {et~al.} 2012, \apj, 744, 44

\bibitem[{{Williams} {et~al.}(2009)}]{williams2009}{Williams}, R.~J., {Quadri}, R.~F., {Franx}, M., {van Dokkum}, P., \& {Labb{\'e}}, I.  2009, \apj, 691, 1879

\end{thebibliography}
\end{document}